\documentstyle[aps,prb]{revtex}

\def\be{\begin{equation}}
\def\fe{\end{equation}}
\def\bea{\begin{eqnarray}}
\def\fea{\end{eqnarray}}
\def\ns{{n_s}}
\def\ninf{{n_\infty}}
\def\Ominf{{\Omega_\infty}}
\def\Binf{{B_\infty}}
\def\g{{g_c}}
\def\v{{\cal V}}
\def\B{{\cal B}}
\def\E{{\cal E}}
\def\A{{\cal A}}
\def\kin{{\text{kin}}}
\def\mag{{\text{mag}}}
\def\grad{{\text{grad}}}
\def\con{{\text{con}}}
\def\Hc{{H_c}}
\def\r{r}

\begin{document}
\draft
\title{Bogomol'nyi Limit For Magnetic Vortices In Rotating
Superconductor} 

\author {Brandon Carter, David Langlois, Reinhard Prix}
\address{D\'epartement d'Astrophysique Relativiste et de Cosmologie,\\
Centre National de la Recherche Scientifique, \\
Observatoire de Paris, 92195 Meudon, France.}
\date{\it August 1999}

\maketitle 

\begin{abstract}
{\bf Abstract.} 
This work is the sequel of a previous investigation of  stationary and
cylindrically symmetric vortex configurations for simple models
representing an incompressible non-relativistic superconductor in a
rigidly rotating background. In the present paper, we carry out  our
analysis  with  a generalized Ginzburg-Landau description of the
superconductor,  which provides a prescription for the radial profile
of the normal density within the  vortex. Within this framework,  it
is shown that the Bogomol'nyi limit  condition marking the boundary
between type I and type II behavior is  unaffected by the rotation of
the background. 
\end{abstract}

\section{Introduction}

The present work is the sequel of a previous investigation \cite{cpl98} 
on the energy of vortices in a rotating superconductor, where it was 
shown in particular that the magnetic and kinetic contributions to the 
energy density that are proportional to the background angular velocity 
remarkably cancel. The motivation which brought us to study vortices
in rotating superconductors is the study of the interior of neutron stars, 
more specifically their inner core which is believed to contain a proton 
superconductor. Although we worked out the macroscopic description of an 
array of magnetic vortices and superfluid vortices in a general
relativistic framework \cite{cl98} necessary for refined analysis of
neutron stars, thus generalizing
the earlier work of Lindblom and Mendell \cite{ml91} 
in a Newtonian approach, we 
have here restricted our analysis to a Newtonian framework for
simplicity.  

In our  previous work,\cite{cpl98} we  left undetermined 
the explicit structure of the vortex core. One of the purposes of the 
present work is to provide a specification for  the profile of the 
condensate particle density, based on a Ginzburg-Landau type approach.
Our treatment will however not be restricted just to the widely used 
standard Ginzburg-Landau description, but will also be valid for a
generalized version (more readily justifiable by heuristic
considerations\cite{Tin65,leggett}), leaving arbitrary the coupling
constant $\g$ that enters the gradient energy density.

The second purpose of this work is to reexamine the question of the 
Bogomol'nyi limit \cite{Bogo76} 
in the context of a superconductor in a rotating background.
The conclusion will be that the usual boundary between type I and type II
superconductors remains unaffected by the rotation of the background. 

Before entering the details of this work, let us recall the essential
features of the  model and define the relevant quantities.
The superconducting matter will consist of 
a charged superfluid component,  represented by
a locally variable number density $\ns$ of bosonic particles 
characterized by an effective mass $m$, a charge $q$, and of an ordinary 
 component of opposite charge which locally compensates the charge 
of the first component. The essential property 
that distinguishes the superfluid constituent
from ordinary matter is that its momentum is directly related to  the
phase variable $\varphi$ (a scalar with period $2\pi$) of the boson
condensate, according to the expression  
\be 
m \vec{v} + q \vec{A} = \hbar\vec\nabla \varphi\,.\label{0} 
\fe 
In this formula, $\hbar$ is the Dirac-Planck constant, $\vec{A}$ is the
magnetic vector potential and $\vec{v}$ the velocity of the Bose
condensate.
The whole system will be described, in addition to the relation (\ref{0}),
by the Maxwell equations,
\be \vec\nabla \times \vec{B} = 4\pi \vec{\j} \,,\label{equMaxwell}\fe
where $\vec{B}$ is the magnetic field, related to the magnetic potential 
vector by the usual relation $\vec{B}= \vec\nabla \times \vec{A}$, and 
$\vec{\j}$ is the electric current, which consists of the sum of the
currents due to the condensate component and to the ordinary component. 
The ordinary component current will be supposed to be that of a rigidly 
rotating fluid. To be able to solve the coupled system of equations, one
needs a prescription concerning the spatial evolution of the 
condensate particle number density. It will be given by an energy 
minimization principle, using the energy functional corresponding 
to a generalized Ginzburg-Landau approach.

The plan of the paper will be the following. In section
\ref{secEquations}, we shall use the cylindrical symmetry and
introduce new variables to simplify the system of coupled
equations. Section \ref{secEnergy} will be devoted to the energy
minimization principle, which will give a prescription for the
determination of $\ns$. 
And finally, section \ref{secBogomolnyi} will deal with the Bogomol'nyi
limit condition.

\section{System of coupled equations}
\label{secEquations}

The scenarios we shall consider will be of the usual kind, in which each
individual vortex is treated  as  a stationary, cylindrically symmetric
configuration consisting of a rigidly rotating background medium with
uniform angular velocity $\Ominf$, say, together with a charged
superfluid constituent  in a state of differential rotation with a velocity
$v$, which tends at large distance towards the rigid rotation value given by
$\Ominf \r$, where $\r$ is the cylindrical radial distance from the
axis. It will be supposed that the superfluid particle 
number density $\ns$ is a monotonically
increasing function of the cylindrical radius variable $\r$, tending 
asymptotically 
to a constant value, $\ninf$, say, at large distances from the axis.
 It will be supposed that the local charge
density is canceled by the background so that there is no electric field,
but that there is a magnetic induction field with magnitude $B$ and
direction parallel to the axis, whose source is the axially oriented
electromagnetic current whose magnitude $j$ will be given by 
\be j=q\ns(v-\Ominf \r)\,.\label{5}\fe 
The relevant Maxwellian source equation for the magnetic field
(\ref{equMaxwell}) will have the familiar form
\be {dB\over d\r}= -4\pi j\, .\label{4}\fe
The other relevant Maxwellian equation is the one governing
the axial component $A$
(which in an appropriate gauge will be the only component) of the 
electromagnetic potential covector, which will be related to the
magnetic induction by
\be {d(\r A)\over d\r}=\r B\, .\label{3}\fe
The essential property distinguishing the superconducting case from
its  ``normal'' analogue is the London flux quantization condition
(\ref{0}), which in the present context (where all physically relevant
quantities  depend only on the cylindrical radius $\r$)
will be expressible in the well known form~\cite{Tilley}
\be mv+qA={N\hbar\over\r}\, ,\label{2}\fe
where  $N$ is the  phase winding number, which must be an integer.

Before proceeding, it will be useful to take advantage of the
possibility of transforming the preceding
system of equations to a form that is not just linear but also homogeneous,
by replacing the variables $v$, $B$, $A$ by corresponding variables
$\v$, $\B$, $\A$ that are defined by
\be \v=v-\Ominf\r\, , \label{11} \fe
\be \B= B-\Binf\, ,\label{12}\fe
\be \A=A- {1\over2}\r \Binf\,,\label{12a}\fe
where $\Binf$ is the uniform background magnetic field
value that would be generated by a  rigidly  rotating superconductor,
which is  given by the London formula
\be \Binf= -{2m\over q}\Ominf \,, \label{12b}
\fe
obtained by combining (\ref{2}) and (\ref{3}) in the special case of rigid
corotation, i.e., with $v = \Ominf\r$. 

In terms of these new variables the equation (\ref{3}) will be 
transformed into the form
\be {d(\r \A)\over d\r}=\r \B\, ,\label{3c}\fe
while the other differential equation (\ref{4}) will be
transformed into the form
\be {d\B\over d\r}=-4\pi j\, ,\label{13}\fe
in which, rewriting equation (\ref{5}), we shall have
\be j= q\ns\v\, .\label{2b}\fe
Finally, the flux quantization condition (\ref{2}) will be converted into the 
form
\be m\v+q\A={N\hbar\over\r}\, ,\label{2c}\fe
which can be used to transform (\ref{3c}) into
\be 
{m\over q\r}{d (\v\r)\over d\r}=-\B\, .\label{14}
\fe
The advantage of this reformulation is that unlike $v$, $B$, and $A$,
the new variables $\v$, $\B$ and $\A$ are subject just to
homogeneous boundary conditions, which are simply that they all
tend to zero as $\r\rightarrow\infty$.

\section{Energy minimization principle} 
\label{secEnergy}

The  equations of the previous section 
are not sufficient by themselves to fully determine 
the system.
In order to specify the radial distribution of the
condensate particle number density $\ns$ we will  use an energy
minimization principle based on a model in
which the condensate energy density is postulated to
be given as the sum of a gradient contribution
and a potential energy contribution by an expression of the form
\be \E_\con= \E_\grad+ V\,,\label{A1} \fe
where the contribution $\E_\grad$ is proportional to the
square of the gradient of $\ns$ with a coefficient that, like the
potential energy contribution $V$, is given as an algebraic function of
$\ns$ by some appropriate ansatz. The use of such a model
as a fairly plausible approximation is justifiable by heuristic
considerations \cite{Tin65} that motivate the use of an ansatz of
what we shall refer to as the Ginzburg type, according to which the
energy contribution is postulated to have the form
\be \E_\grad=
{\g^2\hbar^2\over 8 m\,\ns}\left({d\ns\over d\r}\right)^2   
\, ,\label{A2}\fe
where $\g$ is a dimensionless  coupling constant,
while the potential energy density $V$ is given in terms of some constant
proportionality factor $\E_c$, say, by the formula
\be V=\E_c\left(1-{\ns\over\ninf}\right)^2 \,,\label{A3}\fe
which provides a particularly convenient ansatz for interpolation in 
the theoretically intractable intermediate region between the 
comparatively well understood end points of the allowed 
range $0\leq\ns\leq\ninf $.
The constant $\E_c$ is interpretable as representing
the maximum condensation energy density.
Its value is commonly 
expressed in terms of the corresponding critical value $\Hc$, say,
representing the strength of the maximum magnetic field that can be expelled
from the superconductor by the Meissner effect, to which it is evidently
related by the formula
\be \E_c={\Hc^2\over 8\pi}\, .\label{A4}\fe 
The total energy density associated with a vortex will be of the general 
form 
\be \E= \E_\mag+\E_\kin
+\E_\con\, ,\label{95}\fe
where $\E_\mag$, $\E_\kin$ and 
$\E_\con$
are respectively the magnetic, kinetic and  condensate energy contributions.
More precisely,  $\E_\mag$ is the extra magnetic energy density 
arising from a non-zero value of the phase winding number $N$,
i.e., the local deviation from the magnetic energy density due just to
the uniform field $\Binf$ (associated with of the state of rigid
corotation characterized by the given background angular velocity
$\Ominf$), namely,
\be \E_\mag=
{B^2\over 8\pi}-{\Binf^2\over 8\pi}
\, ,\label{32}\fe
while $\E_\kin$ is the corresponding deviation of the kinetic
energy from that of the state of rigid corotation characterized by the given
background angular velocity $\Ominf$, namely,
\be \E_\kin=
{m\over 2}\ns\left(v^2-\Ominf^2\r^2\right)
\, . \label{33}\fe
It is convenient for many purposes to express such a model in terms
of a dimensionless amplitude $\psi$
that varies in the range $0\leq\psi\leq 1$ according to the 
conventional specification
\be \ns=\psi^2 \ninf\, .\label{22}\fe
Within the general category of Ginzburg type models  as thus
described,  the special case of the standard kind of Ginzburg-Landau
model is  characterized more specifically by the postulate that the
gradient coupling constant $\g$ should be exactly equal to unity. This
ansatz has the attractive feature of allowing the theory to be neatly 
reformulated in terms of a complex variable $\Psi\equiv\psi{\rm e}^{i\varphi}$ 
where $\varphi$ is the phase that appears in (\ref{0}) in a manner
that is evocative of the Schroedinger model for a single particle.
Indeed, it is easy to verify that in the case of $\g=1$, the gradient
term (\ref{A2}) and the kinetic term in (\ref{33}) can be rewritten,
using (\ref{0}) and (\ref{22}), to give the usual
Ginzburg--Landau type gradient term, i.e.,
\be
{\hbar^2\over 8m\ns}\left(\vec\nabla \ns\right)^2 + 
{m\over2}\ns {\vec{v}}^{\,2} = 
{\hbar^2\ninf\over 2m} 
\left\vert \vec{\cal D}\Psi\right\vert^2\,,
\fe
where the covariant derivative is defined as
\be
\vec{\cal D} \equiv \vec\nabla - {iq\over\hbar}\vec A\,.
\fe
However, although there are physical reasons~\cite{Tin65} for expecting
that $\g$ should be comparable with unity, the seductive supposition that
it should exactly satisfy the Landau condition
$\g=1$ is more dubious.\cite{leggett}  This more specialized ansatz will not be
needed for the work that follows, which applies to the generalized
Ginzburg  category with no restriction on the parameter $\g$.

Using the expression 
\be 
\E_\mag+\E_\kin
= {\B^2\over8\pi}+{m\over 2q}j\v+{\Binf\over8\pi\r}
{d\over d\r}\left(\r^2\B\right) \label{47c}\fe
for the first two terms in the
combination (\ref{95}), it can be seen that the total energy 
density arising from the presence of the vortex will be given by 
\be \E
= {\B^2\over 8\pi}+{\Hc^2\over 8\pi}\left(1-\psi^2\right)^2
+{\ninf\over 2 m}\left[\left(\g\hbar{d\psi\over d\r}\right)^2+
\left(m\v\psi\right)^2\right]
+ {\Binf\over 8\pi\r} {d(\r^2\B)\over d\r}\, .\label{A6}\fe
The equation governing the distribution of the condensate particle
number density $\ns$ is obtained by requiring
that the integral of the energy density (\ref{A6}) be stationary with respect
to variation of $\ns$ or equivalently of $\psi$, which gives the
field equation for the latter in the form
\be {\g^2\hbar^2\over m \r}{d\over d\r}\Big(\r{d\psi\over d\r}\Big)=
m\v^2\psi-{4\over \ninf}\E_c\psi(1-\psi^2)
\, .\label{A7}\fe
When this equation is satisfied, it can be seen that the energy density
(\ref{A6}) will reduce to a value given by
\be \E
={\B^2\over 8\pi}+{\Hc^2\over 8\pi}\left(1-\psi^4\right)
+ {1\over\r}{d\over d\r}\left( \Binf{\r^2\B\over 8\pi}
+\r{\g^2\hbar^2\ninf\over 4 m}{d\psi^2\over d\r}\right) \, ,\label{A8}\fe
in which the last term is a divergence that goes out when integrated,
so that for the total energy per unit length one is left simply with
\be U={1\over 8\pi}\int\left(\B^2
+\Hc^2\left(1-\psi^4\right)\right)\, dS
\, .\label{A8b}\fe

\section{Bogomol'nyi inequality}
\label{secBogomolnyi}

We shall now try to rewrite the energy density associated with 
the vortex in a different form.
Let us begin by writing the relation
\be
\Big(\g\hbar{d\psi\over d\r}\Big)^2 + \big( m\v\psi\big)^2
=\left(\g\hbar{d\psi\over d\r} \mp  m\v\psi\right)^2 
\pm\g\hbar q\psi^2\B
\pm{\g\hbar m\over q\ninf}{d (\r j)\over\r d\r}\, ,\label{A9}
\fe
which can be obtained by rewriting $\psi (d\psi /d\r)$ in terms of 
$dj/d\r$ and $d\v/d\r$, the latter term being transformed by use of
the Maxwell equation (\ref{14}). In the first term on the right hand side of 
(\ref{A9}), one takes the minus sign if $\v$ is positive, the plus sign 
otherwise. 
We are thus able to  obtain a Bogomol'nyi type reformulation of (\ref{A6}) 
that is given by
\bea 
\E&=&{\g\hbar q \ninf\over 2 m}
\vert\B\vert+{1\over 2\pi}\left( {\vert\B\vert\over 2}+
{\pi\g\hbar q \ninf\over m}(\psi^2-1)\right)^2
+{\ninf\over 2 m}\left(\g\hbar{d\psi\over d\r} - 
m \vert\v\vert\psi\right)^2 \nonumber\\
&+&\left({\Hc^2\over 8\pi}-{\pi\g^2\hbar^2 q^2 \ninf^2
\over 2 m^2}\right) \big(1-\psi^2\big)^2
+{1\over \r}{d\over d\r}\Big( {\Binf \r^2 \B\over 8\pi}
+{\g\hbar\r\vert j\vert\over 2 q}\Big)\, ,\label{A10}
\fea
which generalizes the original version\cite{Bogo76} by inclusion
of the term with the coefficient $\Binf$, which
allows for the effect of the background rotation velocity $\Ominf$.
Since this extra term is just the divergence of a quantity that vanishes
both on the axis and in the large distance limit it gives no 
contribution to the corresponding integral expression,
which therefore has the same form as the usual Bogomol'nyi
relation for the non rotating case.\cite{JR79} More generally,
performing the Bogomol'nyi trick (\ref{A9}) just for a fraction
$f$ of the combined kinetic and gradient contribution, one can see 
that for a vortex with (relative) magnetic flux 
\be \Phi=\int \B dS\,, \fe
the total vortex energy per unit length and per unit of flux will be
expressible in the form 
\bea
U &=& {f\Hc\over 4\pi \kappa}\vert\Phi\vert
+{\Hc^2\over 8\pi} 
\Big(1-{f^2\over\kappa^2}\Big)\int \big(1-\psi^2\big)^2\, dS
+{\Hc^2\over 8\pi}\int\left( 
{\vert\B\vert\over \Hc}-
{f(1-\psi^2)\over\kappa}\right)^2 dS \nonumber\\
&+& {f \ninf\over 2 m} \int\Big(\g\hbar{d\psi\over d\r} 
- m\vert\v\vert\psi\Big)^2 dS  + {(1-f)\ninf\over 2 m}\int
\left(\Big(\g\hbar{d\psi\over d\r}\Big)^2+ \big(m\v\psi\big)^2\right)\,dS\, ,\label{A11}
\fea
in which  $\kappa$ is a dimensionless constant given by the definition
\be \kappa
={m \Hc\over 2\g\pi\hbar q \ninf}
\, .\label{A11a}\fe
It can be convenient to introduce the so-called London penetration length 
$\lambda$, defined by the expression
\be \lambda^2={m\over 4\pi q^2 \ninf}\,, \fe
and the usual flux quantum
\be \tilde\Phi={2\pi \hbar\over q}\,. \fe
It is to be observed that all the terms in this expression will be
non-negative provided the quantity $f$ is chosen not only so
as to lie in the range $0\leq f\leq 1$ but also so as to satisfy
$f\leq\kappa$, a requirement that will be more restrictive 
if $\kappa\leq 1$. In the latter case we can maximize the first
term on the right of (\ref{A11})
by choosing $f=\kappa$, thereby incidentally eliminating
the second term , so that we obtain the lower limit
\be \kappa\leq 1\ \ \ \Rightarrow\ \ \
{U\over \vert\Phi\vert}\geq {\Hc\over 4\pi}\, .\label{A12a}\fe
If $\kappa\geq 1$ we shall be able to choose $f=1$, thereby eliminating 
the final term in (\ref{A11}), so that we obtain the inequality
\be 
\kappa\geq 1\ \ \ \Rightarrow\ \ \
{U\over \vert\Phi\vert}\geq{\Hc\over 4\pi\kappa}\, .\label{A12b}
\fe
More particularly it can be seen that choosing $f=1$ will eliminate
both the second term and the last term on the right of (\ref{A11}) 
in the special Bogomol'nyi limit case characterized the condition
\be \kappa=1 \, .\label{A12}\fe
(Readers should be warned that much of the relevant 
literature\cite{Kramer71,JR79} follows a rather awkward tradition in which
the symbol $\kappa$ is used for what in the present notation scheme
would be denoted by $\sqrt{2}\,\kappa$, which instead of (\ref{A12})
makes the critical condition come out to be $\kappa=1/\sqrt{2}\,$.)
In this Bogomol'nyi limit, the energy per unit flux per unit 
length (\ref{A11}) will be minimized by imposing the conditions
\be 
\g\hbar{d\psi\over d\r}= m\vert\v\vert\psi\, ,\label{A14}
\fe
(with the sign adjusted so as to make the right hand side positive),
and
\be \vert\B\vert={\g\tilde\Phi\over 4\pi\lambda^2}
\big(1-\psi^2\big)\, .
\label{A15}\fe
which (as when the background is non-rotating\cite{VS76})
will automatically guarantee the solution of the field equations  
in this case, annihilating the last two terms in (\ref{A11}) so that
one is left simply with
\be
{U\over \vert\Phi\vert}= {\Hc\over 4\pi}\, .\label{A16}
\fe
The qualitative distinction between what are known\cite{Gen65} as Pippard type
or type I superconductors on one hand and as London type or type II
superconductors on the other hand is based on the criterion of whether or not,
for a given total flux, the energy will be minimized by gathering the flux in
a small number of vortices, each with large winding number $N$ or by separating
the flux in a large number of vortices, each just with unit winding
number. Within the framework of our analysis, a model may be
characterized as type I if the derivative with respect to $|\Phi|$ of
$U/|\Phi|$ is always negative, i.e., if $dU/d|\Phi|< U/|\Phi|$, in
which case the vortices will effectively be mutually attractive, and
as being of type II if the derivative with respect to $|\Phi|$ of
$U/|\Phi|$ is always positive, i.e., if $dU/d|\Phi|> U/|\Phi|$, in
which case the vortices will effectively be mutually repulsive. 
One can of course envisage the possibility of models that are intermediate in
the sense of having a sign for the derivative of $U/|\Phi|$ that depends on
$N$, so that the minimum is obtained for some large but finite value of the
winding number. 

What can be seen directly from (\ref{A16}) is that within the category of
models characterized by the Ginzburg Landau ansatz, the special Bogomol'nyi
limit case lies precisely on the boundary between type I and type II,
since it evidently satisfies the exact equality
\be 4\pi \lambda^2 \Hc=\g\tilde\Phi \ \ \ \Rightarrow\ \ \ \
 {dU\over d|\Phi|}={U\over|\Phi|}\, ,\label{A17}\fe
for all values of $|\Phi|$. The implication of our work is that the well 
known conclusion~\cite{Kramer71,JR79} that there is neither repulsion nor
attraction between Ginzburg model vortices in the Bogomol'nyi limit 
case will remain valid even in the presence of a rotating background. (It has 
also been shown to be generalizable to cases where self gravitation
is allowed for in a general relativistic framework~\cite{Linet88,CG88}).

Since the change of variables performed in section \ref{secEquations}
has transformed the system to a representation that is formally
identical to that of the case with a non-rotating background, it can
also be concluded that the conclusions of the pioneering analysis of
Kramer~\cite{Kramer71} will remain valid, i.e., that in the sense of
the preceding paragraph  the system will be of the Pippard kind 
(type I), if $\kappa<1$, i.e.,
\be 4\pi \lambda^2 \Hc<\g\tilde\Phi\ \ \ \ 
\Rightarrow  \ \ \ \ {dU\over d|\Phi|}<{U\over|\Phi|} \,,\label{A18}
\fe 
and that it will be of the London kind (type II), if $\kappa>1$, i.e.,
if 
\be 4\pi\lambda^2 \Hc>\g\tilde\Phi\ \ \ \ 
\Rightarrow \ \ \ \ {dU\over d|\Phi|}>{U\over|\Phi|} \,, \label{A19} \fe
at least so long as the winding number $N$ and the
parameters $\kappa$ and $\g$ are not too far from the neighborhood
of unity, in which the numerical investigations have been
carried out. It does not seem that
the possibility of an intermediate scenario, with $U/|\Phi|$ minimized
by a winding number $N$ that is finite but larger that one,
can occur within the framework of Ginzburg type models
except perhaps for parameter values that are too extreme to
be of likely physical relevance.

\end{document}